\documentclass[fleqn,usenatbib]{mnras}

\usepackage{newtxtext,newtxmath}

\usepackage[T1]{fontenc}

\DeclareRobustCommand{\VAN}[3]{#2}
\let\VANthebibliography\thebibliography
\def\thebibliography{\DeclareRobustCommand{\VAN}[3]{##3}\VANthebibliography}

\usepackage{graphicx}	
\usepackage{amsmath}	
\usepackage{graphicx}	
\usepackage{amsmath}	

\usepackage{amssymb}	
\usepackage{siunitx} 
\usepackage{booktabs}
\usepackage[flushleft]{threeparttable}
\usepackage{float}
\usepackage{cleveref}
\usepackage{pifont}

\defcitealias{Rowan22}{R22}

\input{anc/fpcommands}
\newcommand{\PHOEBE}{\texttt{PHOEBE}}
\newcommand{\Gaia}{\textit{Gaia}}
\newcommand{\teffratio}{$T_{\rm{eff},2}/T_{\rm{eff},1}$}
\newcommand{\gof}{{\tt goodness\_of\_fit}}

\title[Fundamental Stellar Parameters]{The Value-Added Catalog of ASAS-SN Eclipsing Binaries III: Masses and Radii of Gaia Spectroscopic Binaries}

\author[D. M. Rowan et al.]{D. M.
Rowan,$^{1,2}$\thanks{E-mail: rowan.90@osu.edu},
T. Jayasinghe$^{1,2,3,4}$,
K. Z. Stanek$^{1,2}$,
C. S. Kochanek$^{1,2}$,
Todd A. Thompson$^{1,2,5}$,
\newauthor
B. J. Shappee$^{6}$, and
W. Giles$^{7}$
\\
$^{1}$Department of Astronomy, The Ohio State University, 140 West 18th Avenue, Columbus, OH, 43210, USA\\
$^{2}$Center for Cosmology and Astroparticle Physics, The Ohio State University, 191 W. Woodruff Avenue, Columbus, OH, 43210, USA\\
$^{3}$Department of Astronomy,  University of California Berkeley, Berkeley CA 94720, USA\\
$^{4}$NASA Hubble Fellow\\
$^{5}$Department of Physics, The Ohio State University, Columbus, Ohio, 43210, USA\\
$^{6}$Institute for Astronomy, University of Hawaii, 2680 Woodlawn Drive, Honolulu, HI 96822, USA\\
$^{7}$ASC Technology Services, 433 Mendenhall Laboratory 125 South Oval Mall Columbus OH, 43210, USA\\
}
\date{Accepted XXX. Received YYY; in original form ZZZ}

\pubyear{2022}

\begin{document}
\label{firstpage}
\pagerange{\pageref{firstpage}--\pageref{lastpage}}
\maketitle

\begin{abstract}
Masses and radii of stars can be derived by combining eclipsing binary light curves with spectroscopic orbits. In our previous work, we modeled the All-Sky Automated Survey for Supernovae (ASAS-SN) light curves of more than 30,000 detached eclipsing binaries using \PHOEBE{}. Here we combine our results with \nCrossmatch{} double-lined spectroscopic orbits from \Gaia{} Data Release 3. We visually inspect ASAS-SN light curves of double-lined spectroscopic binaries on the lower main sequence and the giant branch, adding \nVI{} binaries to our sample. We find that only \fracPeriodEccMatch{}\% of systems have \Gaia{} periods and eccentricities consistent with the ASAS-SN values. We use {\tt emcee} and \PHOEBE{} to determine masses and radii for a total of \nStarsMassesRadii{} stars with median fractional uncertainties of \medFracErrorMass{}\% and \medFracErrorRadius{}\%, respectively. 
\end{abstract}




\section{Introduction}

Accurate measurements of stellar masses and radii are crucial tests for models of stellar structure and evolution. Stellar models contain empirical prescriptions for effects like mass loss, convective overshoot, mixing, and rotation that alter the stellar properties and need to be accurately calibrated. This needs to be done for stars of different masses, evolutionary states, and metallicities \citep{Andersen91}.

The measured masses and radii of binary stars are also benchmarks for asteroseismology. Solar-like oscillations can be interpreted using scaling relations to measure stellar masses and radii \citep{Kjeldsen95}. These relations are particularly useful for measuring the masses of red giants, since these stars are not well-separated by mass on a color-magnitude diagram, making mass inference from isochrone fitting challenging. \citet{Hekker10} identified oscillations in an eclipsing red giant using Kepler photometry. Dynamical masses and radii were derived from spectroscopic followup \citep{Frandsen13} and were found to be in agreement with the astereoseismic masses and radii, supporting the scaling relations \citep{Themebl18}. Since then, a number of authors have used Kepler and TESS photometry to identify oscillating giants in eclipsing binaries \citep{Gaulme13, Gaulme14, Beck14, Brogaard18, Benbakoura21, Beck22}, and these systems suggest that radii are overestimated by $\sim5\%$ and masses are overstimated by $\sim15\%$ when using asteroseismic scaling relations \citep{Gaulme16}. 

Accurate stellar parameters are also needed to characterize exoplanets, since most of the observed properties of transiting exoplanets are measured relative to that of their host star \citep{Eastman13, Rodriguez22}. Theoretical evolutionary tracks or empirical relations derived from eclipsing binaries can be used, but this assumes that the star is typical in terms of mass, metallicity, and rotation rate within the sample of stars used to derive these relations \citep{Enoch10, Torres10, Duck22}.

Masses and radii of stars can be determined by starting from catalogs of eclipsing binaries found by photometric surveys such as the Optical Gravitational Lensing Experiment \citep[OGLE,][]{Graczyk11, Pawlak13, Pietrukowicz13, Soszynski16, Bodi21}, Kepler \citep{Prsa11, Slawson11, Kirk16}, the Wide-field Infrared Survey Explorer \citep[WISE,][]{Petrosky21}, the All-Sky Automated Survey \citep[ASAS,][]{Pojmanski02, Paczynski06}, and the Transiting Exoplanet Survey Satellite \citep[TESS,][]{Ricker15,Prsa22}. Physical masses and radii can be determined by combining the eclipsing binary light curve with radial velocity observations \citep[e.g.,][]{Ratajczak21, Helminiak21}.

Large spectroscopic surveys such as the Apache Point Observatory Galactic Evolution Experiment \citep[APOGEE,][]{Majewski17}, the Large Sky Area Multi-Object Fiber Spectroscopic Telescope \citep[LAMOST,][]{Cui12}, and the Radial Velocity Experiment \citep[RAVE,][]{Steinmentz06}, can also be combined with photometric surveys to constrain stellar parameters. Even for systems with few radial velocity epochs, an eclipsing binary light curve can provide the precise period and ephemeris, so stellar and orbital parameters can often be determined \citep[e.g.,][]{Qian17, Qian18, Hambleton22}. However, only fractions of these catalogs contain the double-lined spectroscopic binary orbits necessary to derive stellar masses and radii \citep[e.g.,][]{Kounkel21}.

\Gaia{} DR3 has significantly expanded the quantity of available spectroscopic data. Nearly 1 million stars have mean RVS spectra \citep{GaiaCollaboration2022}. Although individual epoch radial velocities are only available for $<2000$ RR Lyrae and Cepheid stars, \Gaia{} DR3 includes spectroscopic orbit parameters for more than 181,000 single-lined spectroscopic binaries (SB1s) and more than 5,000 double-lined spectroscopic binaries (SB2s) with $G<12$~mag.

In \citet[][hereafter R22]{Rowan22}, we modeled the light curves of more than 30,000 detached eclipsing binaries from the All-Sky Automated Survey for Supernovae \citep[ASAS-SN,][]{Shappee14, Kochanek17, Jayasinghe19}. We used \PHOEBE{} \citep{Prsa05, Prsa16, Conroy20} to fit the $V$- and $g$-band light curves, producing a catalog of orbital periods, eccentricities, inclinations, effective temperature ratios, and sum of the radii relative to the semimajor axis. By combining our catalog with colors and magnitudes from \Gaia{}, 3-dimensional dust maps of the Milky Way from \texttt{mwdust} \citep{Bovy16, Drimmel03, Marshall06, Green19}, and MIST isochrones and evolutionary tracks \citep{Choi16, Dotter16}, we examined the properties of the systems as a function of their absolute magnitude and evolutionary state. In \citet{Rowan22II}, we characterized the properties of more than 700 binaries with spots, pulsations, and triple/quadruple systems using ASAS-SN and \textit{TESS} \citep{Ricker15, Kunimoto21,Huang20a, Huang20b}.

Here, we combine the \citetalias{Rowan22} ``Value-Added'' catalog of eclipsing binaries with the SB2 orbit solutions from \Gaia{} DR3. In Section \ref{sec:sample}, we cross-match the Value-Added catalog with the catalog of \Gaia{} SB2s. We also visually inspect ASAS-SN light curves for SB2s in sparsely populated areas of the color-magnitude diagram (CMD) to identify more eclipsing SB2s. We then compare the \Gaia{} and ASAS-SN orbital periods and eccentricities to identify systems with reliable \Gaia{} orbital solutions. In Section \ref{sec:models} we use \PHOEBE{} to model the light curve of the detached eclipsing binary with the orbital constraints from \Gaia{}. Finally, we present the distribution of stellar parameters in Section \ref{sec:results}.

\section{Eclipsing Binaries with Spectroscopic Orbits} \label{sec:sample}

\begin{figure}
    \centering
    \includegraphics[width=\linewidth]{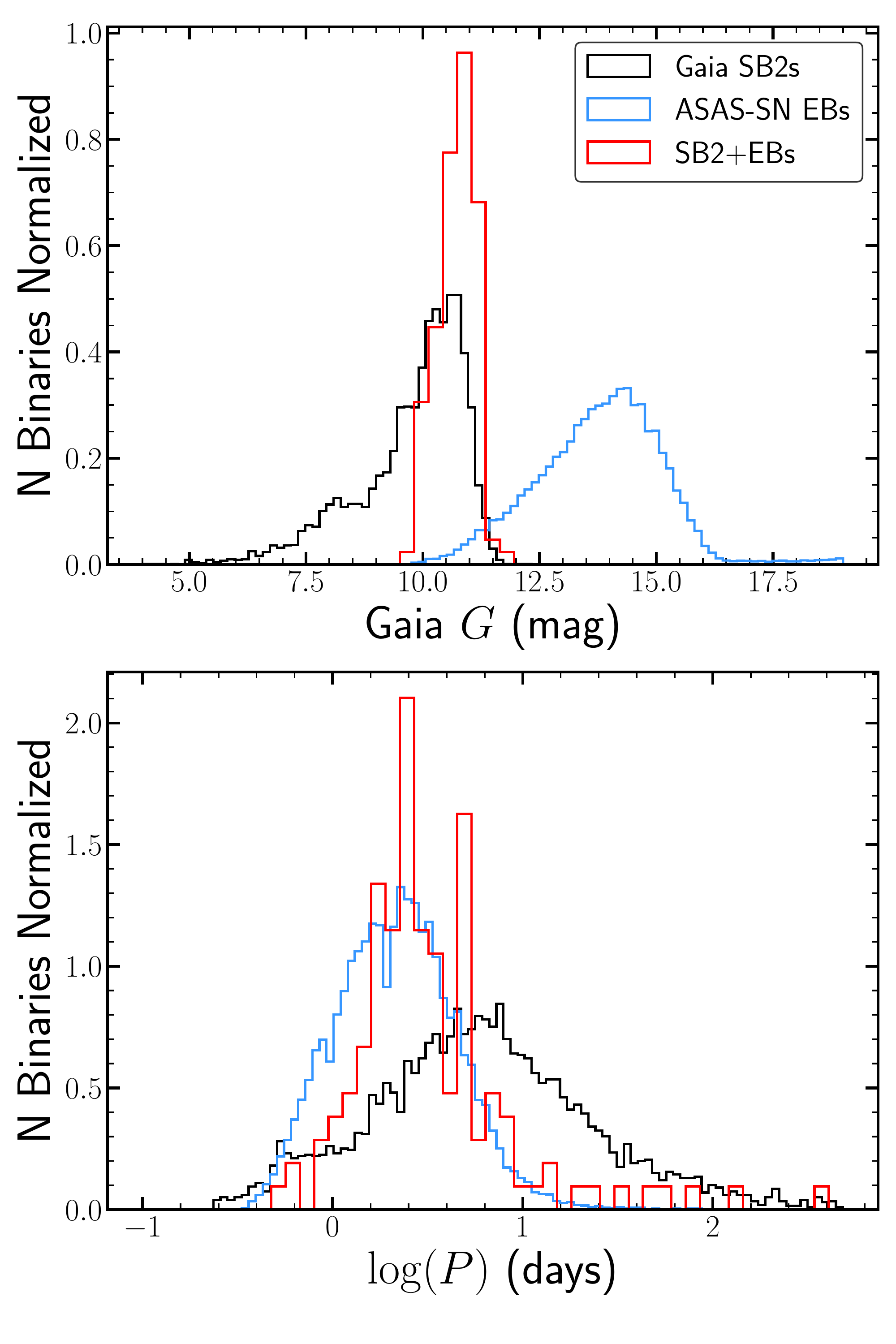}
    \caption{Top: normalized distributions of \Gaia{} apparent $G$-band magnitudes for \Gaia{} SB2s, ASAS-SN eclipsing binaries, and the crossmatched systems. Bottom: distribution of orbital periods. The \Gaia{} spectroscopic binaries extend to longer orbital periods since the detectability of eclipsing binaries drops off as $P^{-2/3}$.}
    \label{fig:parameter_hists}
\end{figure}

\begin{figure*}
    \centering
    \includegraphics[width=1.0\linewidth]{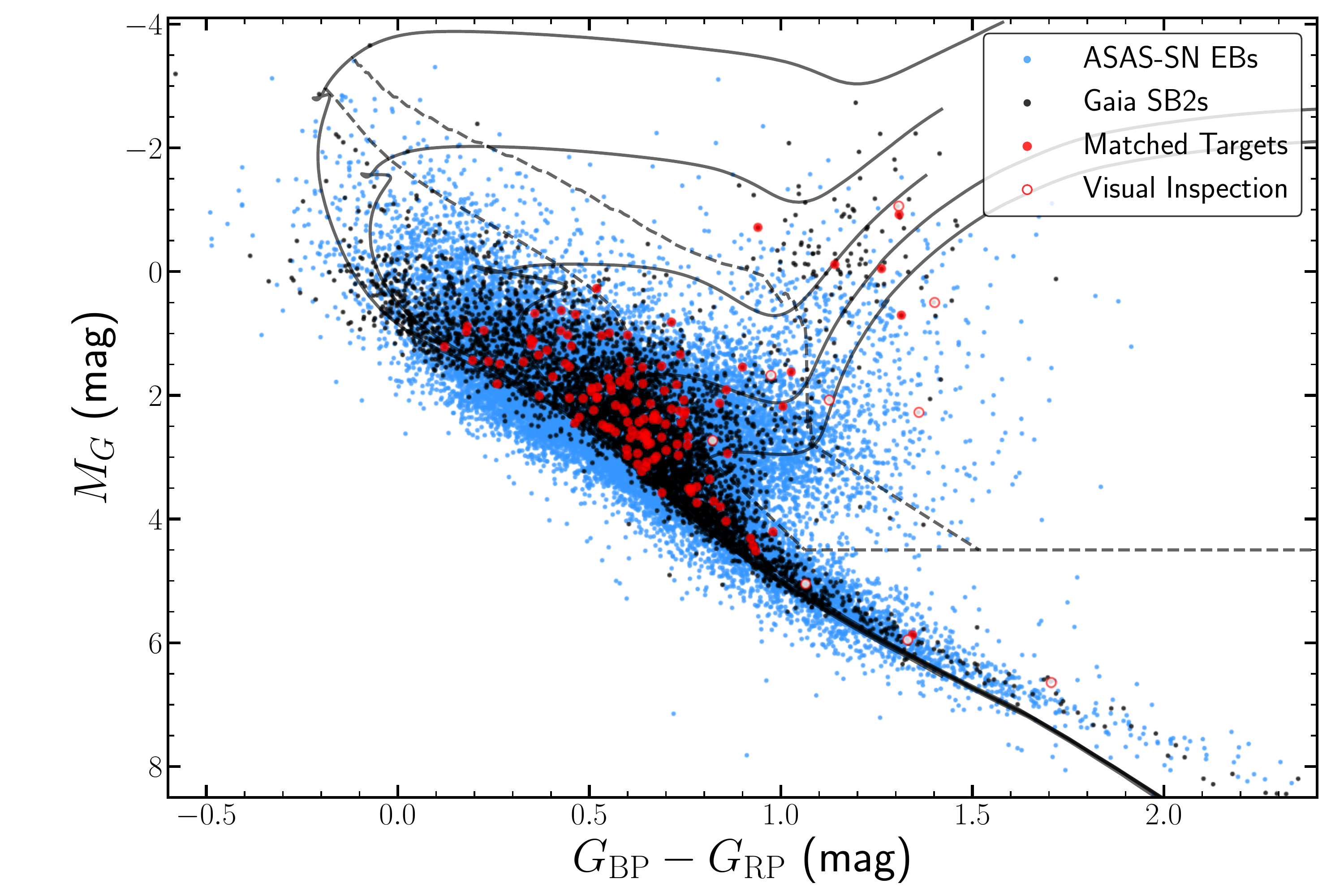}
    \caption{The extinction-corrected \Gaia{} DR3 color-magnitude diagram (CMD). The ASAS-SN detached eclipsing binaries from \citetalias{Rowan22} are shown in blue, the \nSB{} \Gaia{} SB2s are shown in black, and the eclipsing spectroscopic binaries are shown in red. The solid lines show MIST isochrones for ages of $10^8$ to $10^{10}$ years in intervals of $0.5$ dex. The flux of the isochrones are doubled in each band to represent binary stars of equal mass. The dashed lines show the boundaries of the giant and subgiant branches defined by \citetalias{Rowan22}. We note that one binary in the visual inspection group, CM Dra (\Gaia{} DR3 1431176943768690816) is at $M_G\sim10.6$~mag, below the range of this figure.}
    \label{fig:cmd}
\end{figure*}

\begin{figure*}
    \centering
    \includegraphics[width=0.9\linewidth]{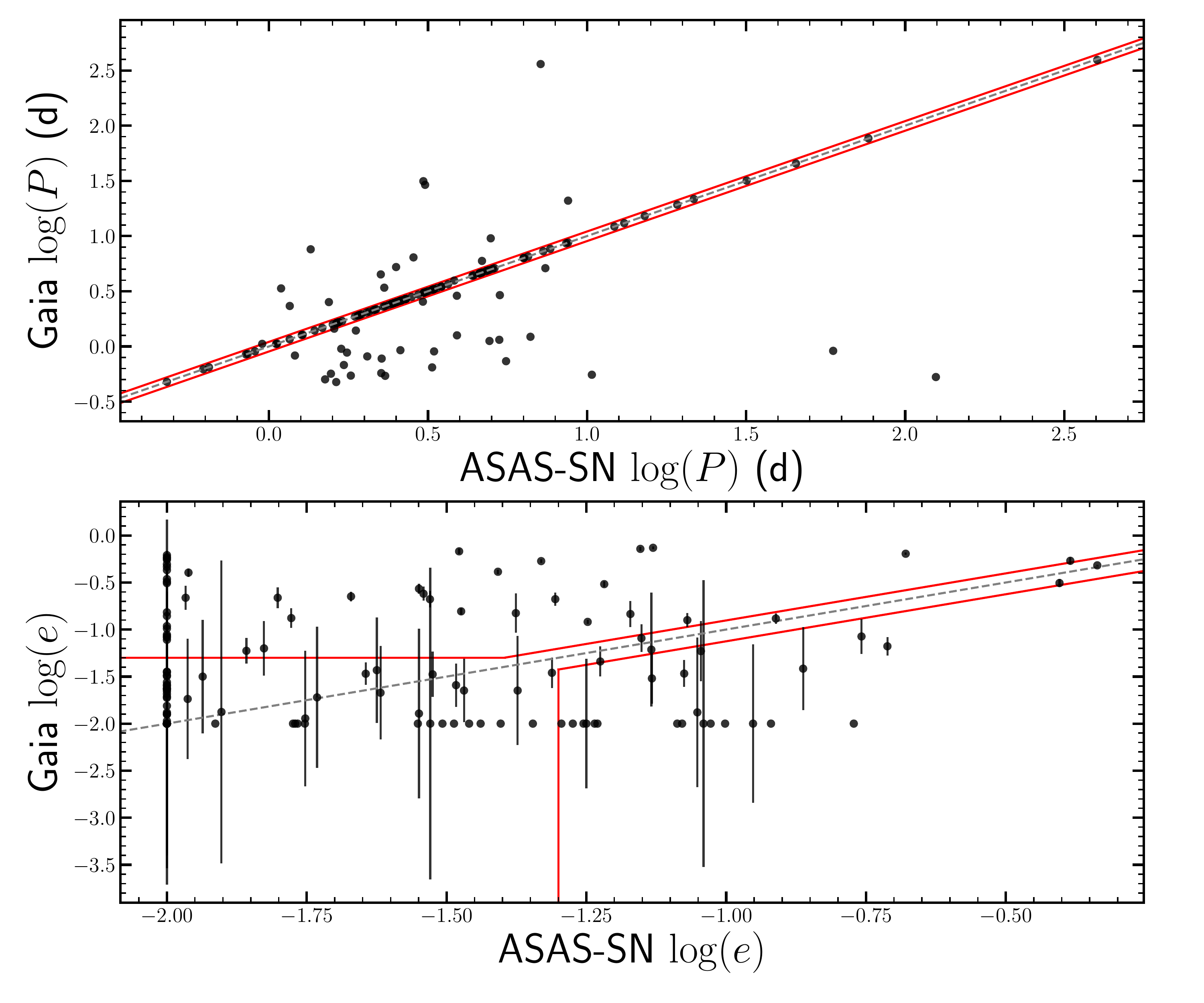}
    \caption{Comparison of periods and eccentricities from the \Gaia{} SB2 orbits and ASAS-SN eclipsing binary light curves. The gray lines show the expected equality between the ASAS-SN and Gaia values. The region in between the red lines shows where we define the periods and eccentricities to be consistent. Eccentricities $\log e< -2 $ are shown as $\log e = -2$ for this figure.}
    \label{fig:period_ecc_compare}
\end{figure*}

\begin{figure}
    \centering
    \includegraphics[width=\linewidth]{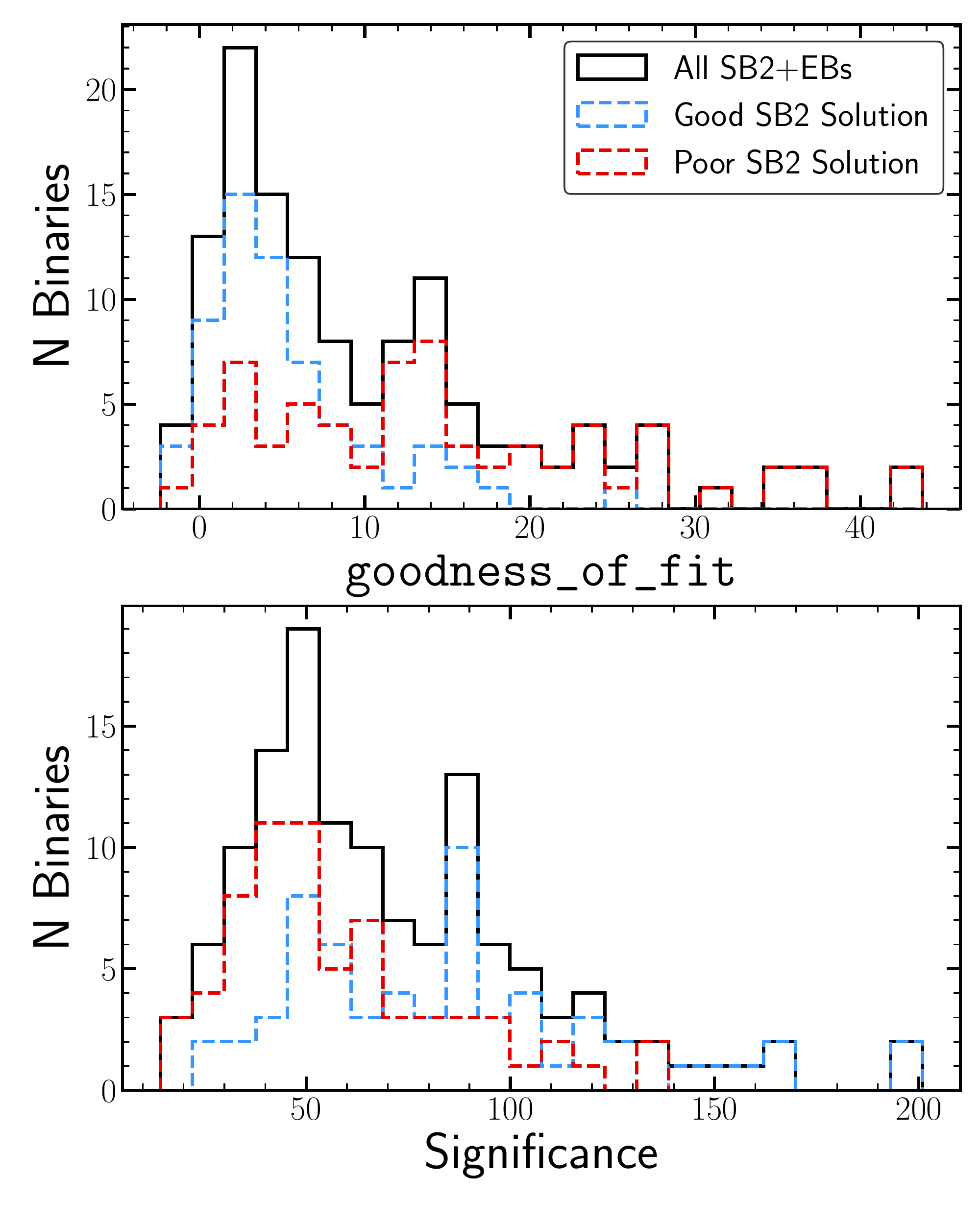}
    \caption{Distributions of the \Gaia{} \gof{} (top) and significance (bottom) for the full SB2+EB sample. We identify good orbital fits by comparing the Gaia and ASAS-SN periods and eccentricities (Figure \ref{fig:period_ecc_compare}).}
    \label{fig:gof_signi_hists}
\end{figure}

\Gaia{} DR3 includes a total of 5,376 systems with SB2 orbit solutions in the \texttt{nss\_two\_body\_orbit} table. The majority of the targets are fairly bright, with apparent $G$ magnitudes ranging from \sbGaiaMin{}~mag to \sbGaiaMax{}~mag and a median of \sbGaiaMed{}~mag. \nSBphotvar{} of the Gaia SB2s are labeled as photometrically variable in Gaia DR3 and \nSBveb{} are included in the {\tt vari\_eclipsing\_binary} table. The optimal magnitude range for ASAS-SN targets is $11 < V <17$~mag and $12<g<18$~mag. Figure \ref{fig:parameter_hists} shows the distribution of apparent $G$ magnitude for the \Gaia{} SB2 catalog and the Value-Added EB catalog.  

We start by performing a positional cross-match with a search radius of $5\arcsec$ and identify \nCrossmatch{} targets in common between the SB2 and EB catalogs. Figure \ref{fig:parameter_hists} shows that these targets are found in the bright tail of the ASAS-SN catalog and the faint end of the SB2 catalog. The bottom panel of Figure \ref{fig:parameter_hists} shows that the majority of systems identified in the cross-match have orbital periods $P<10$~days, although there are a handful of long-period systems included in both catalogs. The orbital period of the \nCrossmatch{} targets ranges from \xmatchPeriodMin{} to \xmatchPeriodMax{}~days with a median value of \xmatchPeriodMed{}~days.

We use distances from \citet{BailerJones21}, which include color and magnitude priors as part of the distance estimate, and extinction estimates from the {\tt mwdust} \citep{Bovy16} 3-dimensional `Combined19' dust map \citep{Drimmel03, Marshall06, Green19} to determine the extinction-corrected absolute magnitude and color. We use the MESA Isochrones \& Stellar Tracks \citep[MIST,][]{Choi16, Dotter16} and follow the procedure described in \citetalias{Rowan22} to divide the CMD into systems with main sequence, subgiant, and giant primaries. To remove systems with poor parallax or extinction estimates, we only report the evolutionary state for systems where the parallax divided by its standard error is $\texttt{parallax\_over\_error} > 10$ and $A_V < 2.0$~mag. Figure \ref{fig:cmd} shows the distribution of the SB2 catalog, the EB catalog, and the crossmatch between them on the CMD. For stars with $M_G\lesssim4$~mag, there are fewer SB2s than ASAS-SN EBs below the binary main sequence since systems of nearly equal mass are more easily characterized in SB2 analysis.

The majority of the SB2+EB binaries are on the main sequence, although \nCrossmatchSG{} and \nCrossmatchRG{} are found on the subgiant and giant branch, respectively. We used ASAS-SN Sky Patrol v2 (Hart~et al., in preparation) to visually inspect the $g$-band light curves of all \nSB{} systems in the \Gaia{} SB2 catalog folded at the \Gaia{} orbital period. In total, we identify 200 additional eclipsing binaries. To expand our sample to stars of a wider range of mass and evolutionary state, we focus on either end of the main sequence ($M_G<1$~mag or $M_G>5$~mag) and stars that are subgiants/giants based on the CMD position and MIST isochrones. Figure \ref{fig:cmd} shows the \nVI{} targets we added to our EB+SB2 catalog to expand our coverage of the CMD. While we add additional low-mass main sequence and giant/subgiant binaries, we find no additional high-mass main sequence binaries that are unsaturated ($V>11$~mag).

Although we expect the \Gaia{} orbital period to be correct for these \nVI{} systems since we identified the eclipses in the phase-folded light curve, we start by running the {\tt astrobase} implementation of Box Least Squares (BLS) periodogram \citep{Bhatti18, Kovacs02} to determine a more precise period. We use a narrow period search window of $\pm20\%$ of the \Gaia{} orbital period. We manually clip outlying points due to saturation effects in the light curves. We then follow the procedure described in \citetalias{Rowan22} to estimate the orbital inclination, eccentricity, argument of periastron, the ratio of effective temperatures, and the sum of the fractional radii with \PHOEBE{} \citep{Prsa05, Prsa16, Conroy20}. To summarize the procedure, we start by combining the results from the \PHOEBE{} geometry estimator \citep{Mowlavi17} and the ``eclipsing binaries via artificial intelligence'' (EBAI) estimator \citep{Prsa08} as initial estimations for the Nelder-Mead optimizer \citep{Gao12}. The resulting model will be used to set the initial conditions for the Markov Chain Monte Carlo (MCMC) fits.

Before we adopt values of the velocity semi-amplitudes from the \Gaia{} SB2 solutions, we compare the ASAS-SN and \Gaia{} periods and eccentricities to identify and remove poor SB2 orbital solutions. Figure \ref{fig:period_ecc_compare} shows the ASAS-SN and \Gaia{} orbital periods and eccentricities. We consider the SB2 solution to be reliable if the \Gaia{} orbital period is within 10\% of the ASAS-SN orbital period. We also require the \Gaia{} eccentricity to be within 25\% of the ASAS-SN eccentricity, or both eccentricities to be $e<0.05$. In total, only \nPeriodEccMatch{} SB2+EB systems meet these quality cuts. \nPeriodBadFaint{} of the systems with discrepant \Gaia{} periods are relatively faint ($G\gtrsim11$~mag), which could suggest lower-quality \Gaia{} SB2 orbital fits in the tail of the magnitude distribution. This magnitude dependence is less clear for the comparison of eccentricities, but some of the systems where the ASAS-SN light curve suggests a near-circular orbit ($e\lesssim0.01$) and \Gaia{} prefers an eccentric orbit are also faint.

\citet{Jayasinghe22} found similar disagreements in period and eccentricity when comparing the \Gaia{} SB1s to the Ninth Catalog of Spectroscopic Orbits \cite[SB9,][]{Pourbaix04} and \citet{Bashi22} used LAMOST and GALAH radial velocities to consider which SB1s may have incorrect orbital parameters. Both found that many of the short-period \Gaia{} SB1s have erroneously high eccentricities. 

Figure \ref{fig:gof_signi_hists} shows the distribution of the \Gaia{} \gof{} and significance parameters. The \gof{} reports the ``Gaussianized Chi-Square'' statistic and is expected to have a median of zero and standard deviation of one, although we note that the median value for the full SB2 sample is \medSBgoodness{}. The top panel of Figure \ref{fig:gof_signi_hists} shows that the SB2s with periods and eccentricities matching the ASAS-SN light curve solution generally have lower \gof{} values, but there are some poor SB2 solutions with $\gof{}<5$. The significance parameter is defined as $K_1/\sigma_{K_1}$. While this statistic is generally useful for rejecting poor (Significance $\lesssim20$) solutions \citep[e.g.,][]{Jayasinghe22}, there does not seem to be a clear value here where \Gaia{} SB2 orbital solutions can be rejected in the absence of constraints from the light curve. \citet{Bashi22} combined the SB1 period, semi-amplitude, number of observations, goodness of fit, RV amplitude, and RV signal-to-noise ratio parameters to construct a ``Score'' statistic, but some of these parameters are not available for the \Gaia{} SB2 sources. Since \Gaia{} DR3 only provides the spectroscopic orbital solutions and not the individual radial velocities, it is clearly important to be cautious when interpreting the ensemble statistics in the \Gaia{} \texttt{nss\_two\_body\_orbit} SB2 tables given that only \fracPeriodEccMatch{}\% of our eclipsing SB2s have reliable \Gaia{} periods and eccentricities.

\section{Eclipsing Binary Model Fitting} \label{sec:models}

\begin{figure*}
    \centering
    \includegraphics[width=\linewidth]{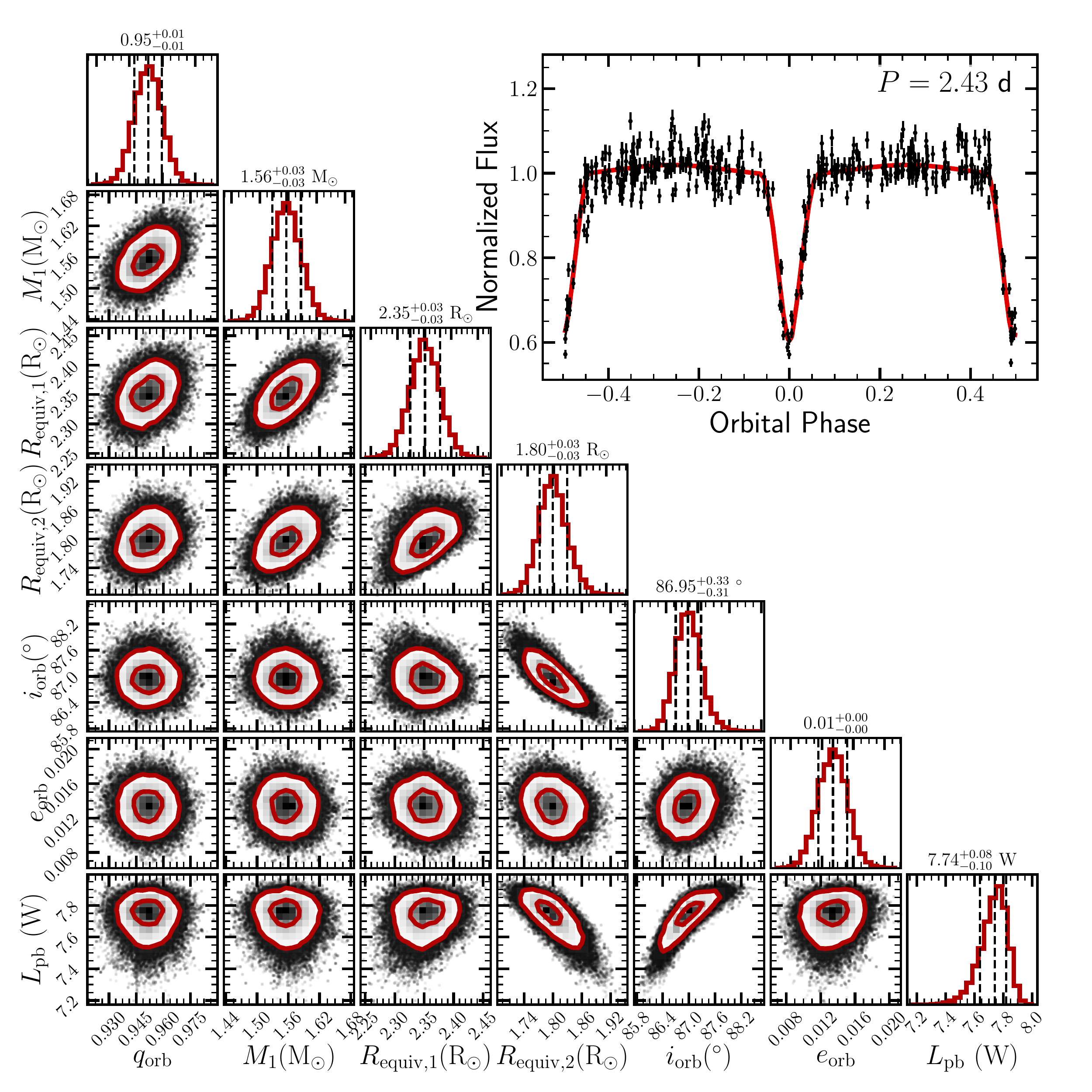}
    \caption{MCMC corner plot and light curve fit for \Gaia{} DR3 154197232963101568.}
    \label{fig:example_mcmc_lc}
\end{figure*}

For the \nPeriodEccMatch{} systems where the period and eccentricity of the \Gaia{} SB2 solution are consistent with the results of the eclipsing binary model fit, we use either the ASAS-SN $V$- or $g$-band light curve, and select the light curve that is less affected by saturation. We change the passband luminosity mode to {\tt component--coupled}, which uses the physical passband luminosity, $L_{\rm{pb}}$ for one star rather than a scaling based on the normalized fluxes. We run 200 iterations of Nelder-Mead optimization with $L_{\rm{pb}}$, the effective temperature ratio \teffratio{}, and the fractional radii $R_1/a$ and $R_2/a$, as free parameters.

We then set the mass ratio to be $q=K_1/K_2$, where $K_1$ and $K_2$ are the \Gaia{} RV semi-amplitudes. We also set the value of the projected semi-major axis of the secondary to
\begin{equation}
    a_2 \sin(i) = K_2 \left(\frac{P}{2\pi}\right)\sqrt{1-e^2}.
\end{equation}

\noindent We use {\tt emcee} \citep{ForemanMackey13} within \PHOEBE{} to perform an MCMC fit on the ASAS-SN light curve.  We set Gaussian priors on $q$ and $a_2\sin(i)$ using the \Gaia{} DR3 values and errors on $K_1$ and $K_2$. The orbital period is fixed at the value from the BLS periodogram of the ASAS-SN light curve.

We sample over the mass ratio, $q$, the primary mass, $M_1$, the radii $R_1$ and $R_2$, the inclination $i$, the eccentricity, $e$, and the passband luminosity $L_{\rm{pb}}$. We use 5000 iterations with a 1000 iteration burn-in and 35 walkers. For some targets where the walkers have not yet converged by 1000 iterations, we run an additional 2000 iterations, increasing the burn-in accordingly. In some cases a few ($\lesssim 5$) of the walkers fail to converge. We manually set cutoffs in the log-probability of the walkers for these systems before adopting the final posterior distributions. Figure \ref{fig:example_mcmc_lc} shows an example of the MCMC posteriors and light curve fit for \Gaia{} DR3 154197232963101568.

We calculate the Roche-lobe filling fraction, $f=R/R_{\rm{roche}}$, where $R_{\rm{roche}}$ can be estimated from the approximation \citep{Eggleton83},
\begin{equation}
    R_{\rm{roche}} / a = \frac{0.49q^{2/3}}{0.6q^{2/3}+\ln{\left(1+q^{1/3}\right)}}.
\end{equation}
The filling fraction can be used to evaluate the degree to which systems are detached and can be thought of as evolving independently, without mass transfer.

\section{Results} \label{sec:results}

\begin{figure*}
    \centering
    \includegraphics[width=0.8\linewidth]{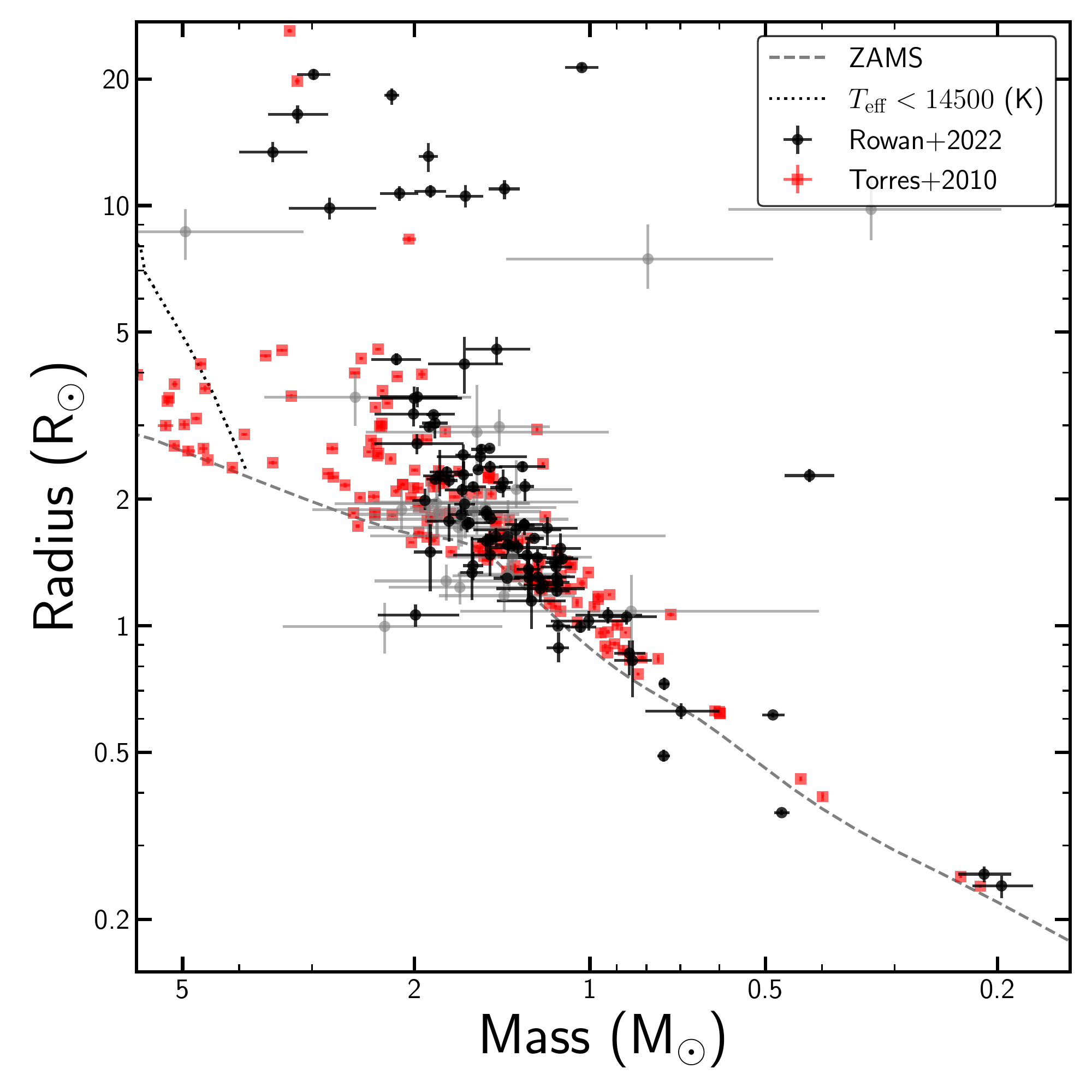}
    \caption{Masses and radii of eclipsing SB2s derived from ASAS-SN eclipsing binary light curves and \Gaia{} SB2 orbital solutions. The gray points shows stars with fractional mass error $>20\%$. The \citet{Torres10} catalog stars are shown in red. The gray dashed line shows the MIST single-star ZAMS isochrone and the gray dotted line shows the cutoff at $T_{\rm{eff}}<14500$~K for the \Gaia{} RVS measurements.}
    \label{fig:mass_radius}
\end{figure*}

\begin{table*}
    \centering
    \caption{Stellar parameters from the MCMC fits to the ASAS-SN light curves with constraints from \Gaia{} spectroscopic orbits. The Roche-lobe filling fractions of the primary and secondary are $f_1$ and $f_2$, respectively. The evolutionary state is based on the CMD position in Figure \ref{fig:cmd}. The full table is available online at \url{https://asas-sn.osu.edu/binaries/mass-radius} and in the electronic version of the paper.}
        \sisetup{table-auto-round,
         group-digits=false}
    \setlength{\tabcolsep}{3pt}
    \renewcommand{\arraystretch}{1.5}
    \begin{center}
        \begin{tabular}{l S[table-format=3.4] r r r r r S[table-format=1.2] S[table-format=1.2] r r r}
\toprule
  {Gaia DR3 Source} &  {Period} &           {Mass Ratio} &                {$M_1$} &                {$M_2$} &                {$R_1$} &                {$R_1$} &  {$f_1$} &  {$f_1$} &                  {$e$} &                   {$i$} & {State} \\
{} & {(d)} & {} & {(M$_\odot$)} & {(M$_\odot$)} & {(R$_\odot$)} & {(R$_\odot$)} & {} & {} & {} & {($^\circ$)} & {}\\ 
\midrule
  40041022325608704 &  2.808646 & $0.99^{+0.03}_{-0.03}$ & $1.28^{+0.11}_{-0.10}$ & $1.27^{+0.13}_{-0.12}$ & $1.47^{+0.11}_{-0.15}$ & $1.30^{+0.15}_{-0.15}$ & 0.338415 & 0.300173 & $0.01^{+0.01}_{-0.01}$ & $85.76^{+0.37}_{-0.21}$ &      MS \\
 154197232963101568 &  2.431772 & $0.95^{+0.01}_{-0.01}$ & $1.56^{+0.03}_{-0.03}$ & $1.48^{+0.04}_{-0.04}$ & $2.35^{+0.03}_{-0.03}$ & $1.80^{+0.03}_{-0.03}$ & 0.556776 & 0.435488 & $0.01^{+0.00}_{-0.00}$ & $86.95^{+0.33}_{-0.31}$ &      SG \\
 428267308105062528 &  3.289680 & $0.95^{+0.21}_{-0.10}$ & $1.89^{+0.85}_{-0.63}$ & $1.83^{+0.60}_{-0.56}$ & $1.95^{+0.27}_{-0.33}$ & $1.96^{+0.43}_{-0.33}$ & 0.354284 & 0.361245 & $0.01^{+0.00}_{-0.00}$ & $83.98^{+0.44}_{-0.33}$ &      MS \\
 453147263375440384 &  4.586150 & $0.98^{+0.02}_{-0.02}$ & $1.84^{+0.10}_{-0.09}$ & $1.81^{+0.12}_{-0.10}$ & $3.03^{+0.10}_{-0.24}$ & $2.28^{+0.35}_{-0.24}$ & 0.445720 & 0.336974 & $0.00^{+0.00}_{-0.00}$ & $84.31^{+0.79}_{-0.55}$ &      MS \\
 533425940213217920 &  4.860321 & $0.97^{+0.02}_{-0.02}$ & $1.54^{+0.06}_{-0.06}$ & $1.48^{+0.08}_{-0.07}$ & $2.63^{+0.07}_{-0.07}$ & $2.39^{+0.08}_{-0.07}$ & 0.394413 & 0.363752 & $0.00^{+0.00}_{-0.00}$ & $87.43^{+0.16}_{-0.13}$ &      SG \\
 535419732749344000 &  2.948269 & $0.87^{+0.32}_{-0.03}$ & $2.44^{+0.43}_{-0.37}$ & $2.13^{+0.33}_{-0.31}$ & $1.91^{+0.12}_{-0.10}$ & $4.33^{+0.22}_{-0.22}$ & 0.340413 & 0.821817 & $0.00^{+0.00}_{-0.00}$ & $87.73^{+1.59}_{-2.20}$ &      SG \\
 555108687465591040 &  6.334426 & $0.95^{+0.14}_{-0.04}$ & $1.00^{+0.17}_{-0.14}$ & $0.96^{+0.14}_{-0.12}$ & $0.91^{+0.11}_{-0.09}$ & $0.95^{+0.08}_{-0.10}$ & 0.131480 & 0.139853 & $0.06^{+0.00}_{-0.00}$ & $88.67^{+0.29}_{-0.15}$ &      MS \\
 583022706417479296 & 12.185384 & $0.96^{+0.01}_{-0.01}$ & $1.20^{+0.03}_{-0.03}$ & $1.15^{+0.04}_{-0.04}$ & $3.93^{+0.52}_{-0.52}$ & $3.93^{+0.73}_{-1.27}$ & 0.347034 & 0.353851 & $0.01^{+0.01}_{-0.01}$ & $83.30^{+4.82}_{-1.34}$ &      SG \\
 681609355666984192 &  2.703459 & $0.96^{+0.14}_{-0.05}$ & $1.54^{+0.29}_{-0.26}$ & $1.48^{+0.23}_{-0.23}$ & $2.52^{+0.17}_{-0.18}$ & $1.47^{+0.21}_{-0.16}$ & 0.559009 & 0.332373 & $0.00^{+0.00}_{-0.00}$ & $79.89^{+0.94}_{-0.94}$ &      MS \\
 690638579515175552 &  2.380733 & $0.96^{+0.01}_{-0.01}$ & $1.18^{+0.03}_{-0.03}$ & $1.13^{+0.04}_{-0.03}$ & $1.62^{+0.21}_{-0.22}$ & $1.67^{+0.14}_{-0.24}$ & 0.428208 & 0.447758 & $0.00^{+0.00}_{-0.00}$ & $85.24^{+0.75}_{-0.38}$ &      MS \\
 690885522954174080 &  0.647598 & $0.81^{+0.04}_{-0.04}$ & $0.87^{+0.10}_{-0.10}$ & $0.70^{+0.11}_{-0.10}$ & $1.05^{+0.04}_{-0.04}$ & $0.63^{+0.03}_{-0.03}$ & 0.721960 & 0.474742 & $0.11^{+0.01}_{-0.01}$ & $84.77^{+0.36}_{-0.62}$ &      MS \\
 947509303493752832 &  3.481821 & $0.99^{+0.05}_{-0.05}$ & $1.23^{+0.20}_{-0.17}$ & $1.22^{+0.23}_{-0.20}$ & $1.31^{+0.10}_{-0.09}$ & $1.23^{+0.09}_{-0.09}$ & 0.263934 & 0.248808 & $0.00^{+0.00}_{-0.00}$ & $88.55^{+0.19}_{-0.15}$ &      MS \\
1204875013864088576 &  3.288118 & $0.97^{+0.12}_{-0.14}$ & $1.64^{+0.72}_{-0.59}$ & $1.56^{+0.86}_{-0.63}$ & $1.97^{+0.38}_{-0.36}$ & $2.89^{+0.85}_{-1.24}$ & 0.375038 & 0.561835 & $0.01^{+0.01}_{-0.01}$ & $72.44^{+4.40}_{-2.17}$ &      MS \\
1272930282497618048 &  5.117104 & $0.95^{+0.07}_{-0.08}$ & $1.43^{+0.40}_{-0.26}$ & $1.34^{+0.44}_{-0.26}$ & $2.98^{+0.29}_{-0.30}$ & $2.11^{+0.36}_{-0.20}$ & 0.440978 & 0.322579 & $0.00^{+0.00}_{-0.00}$ & $88.73^{+0.85}_{-3.41}$ &      SG \\
1387290280744129152 &  0.846993 & $0.76^{+0.62}_{-0.04}$ & $1.77^{+0.58}_{-0.49}$ & $1.36^{+0.40}_{-0.37}$ & $1.28^{+0.12}_{-0.13}$ & $1.46^{+0.14}_{-0.15}$ & 0.577313 & 0.741876 & $0.00^{+0.00}_{-0.00}$ & $87.25^{+0.26}_{-0.21}$ &      MS \\
\bottomrule
\end{tabular}

    \end{center}
    \label{tab:fp_table}
\end{table*}

\begin{figure*}
    \centering
    \includegraphics[width=\linewidth]{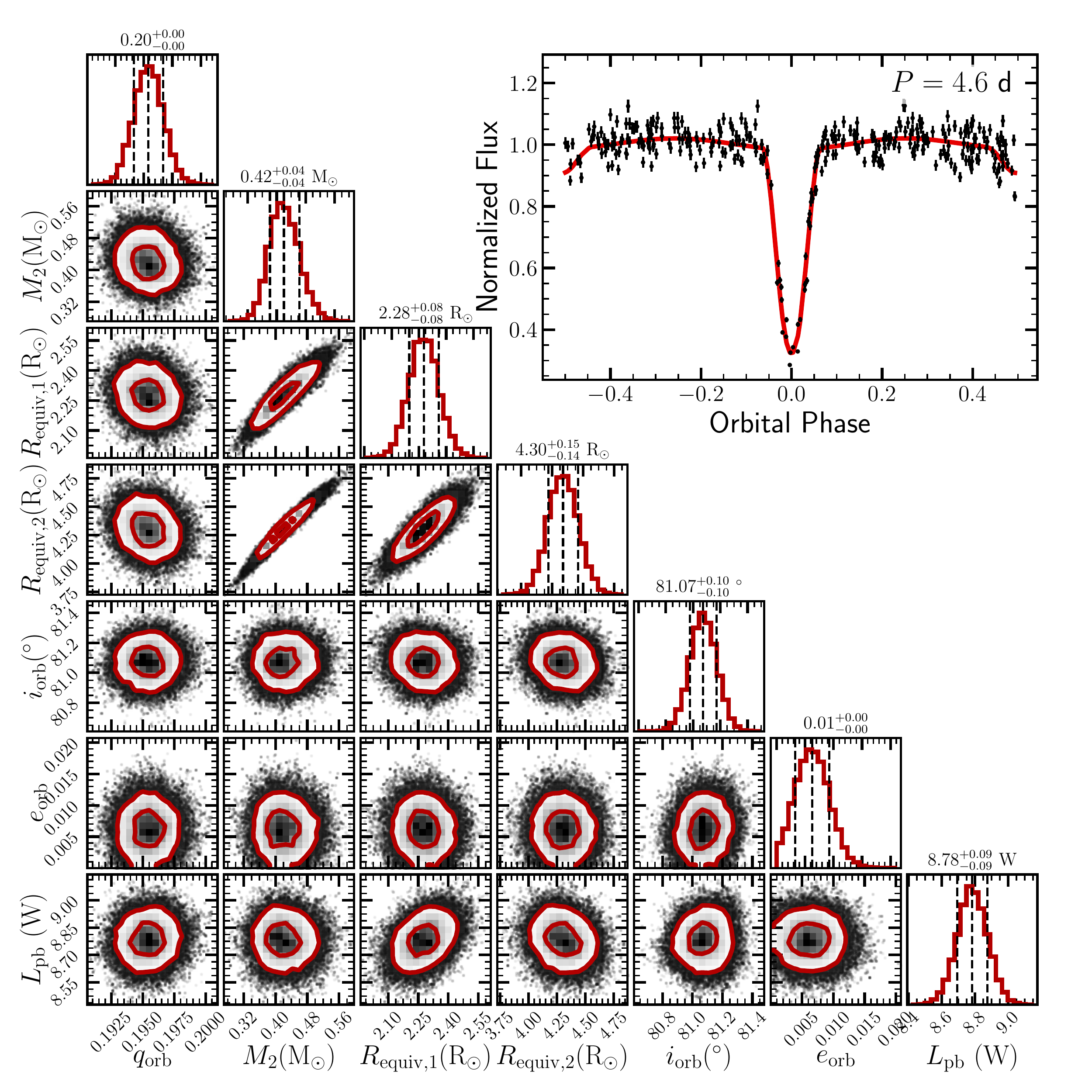}
    \caption{MCMC forner plot and light curve fit for \Gaia{} DR3 1762094209603752192. This binary has one star with $M < 1$~M$_\odot$ and $R>2$~$R_\odot$ and one on the subgiant branch, which could suggest a history of mass transfer.}
    \label{fig:mcmc_lc_176}
\end{figure*}

\begin{figure}
    \centering
    \includegraphics[width=\linewidth]{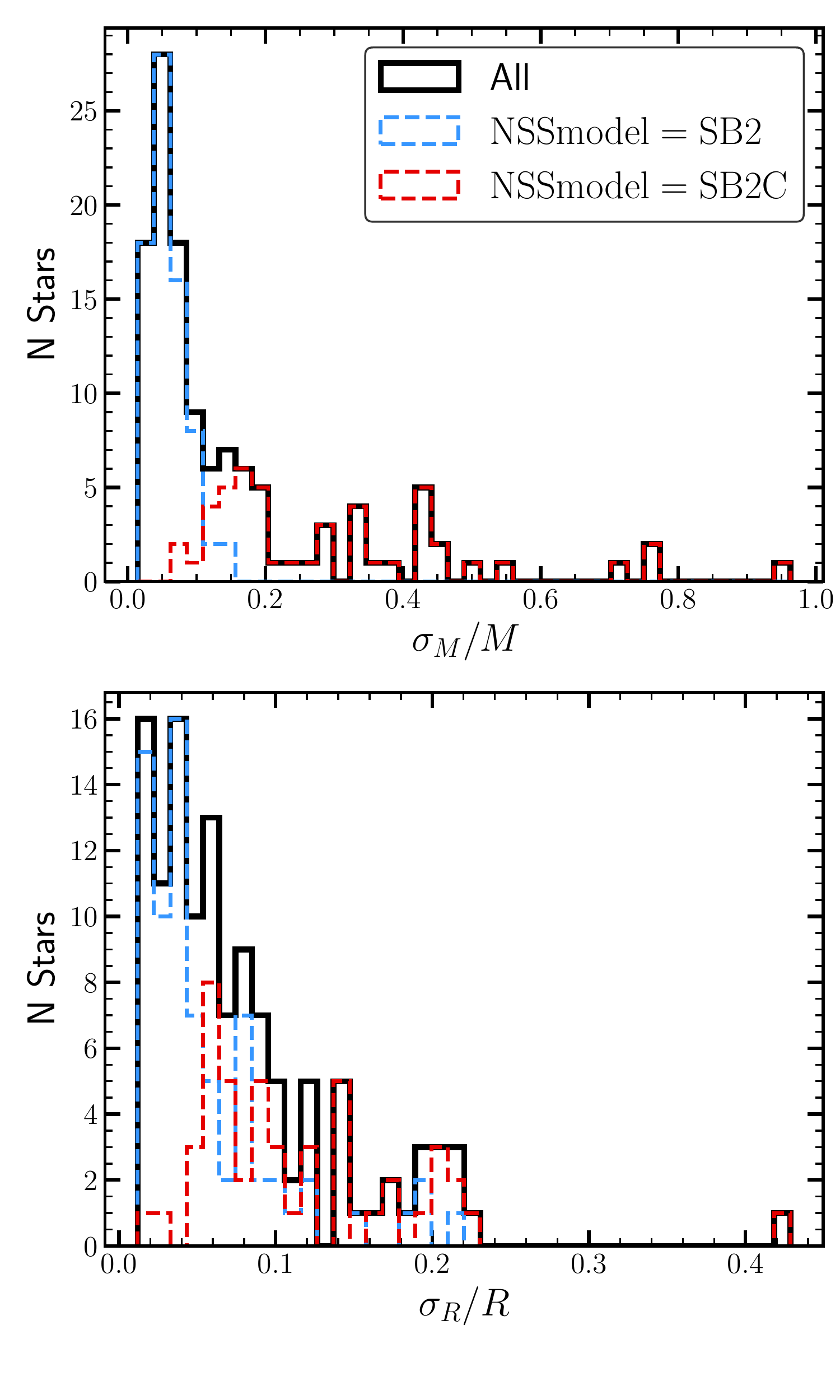}
    \caption{Fractional error on the masses (top) and radii (bottom) from the MCMC fits to the ASAS-SN light curves with orbital constraints from \Gaia{} SB2s. The binaries with the circular orbital solution (NSSmodel=SB2C) make up the majority of the high fractional error tails in mass and radius.}
    \label{fig:frac_error_hists}
\end{figure}

\begin{figure}
    \centering
    \includegraphics[width=\linewidth]{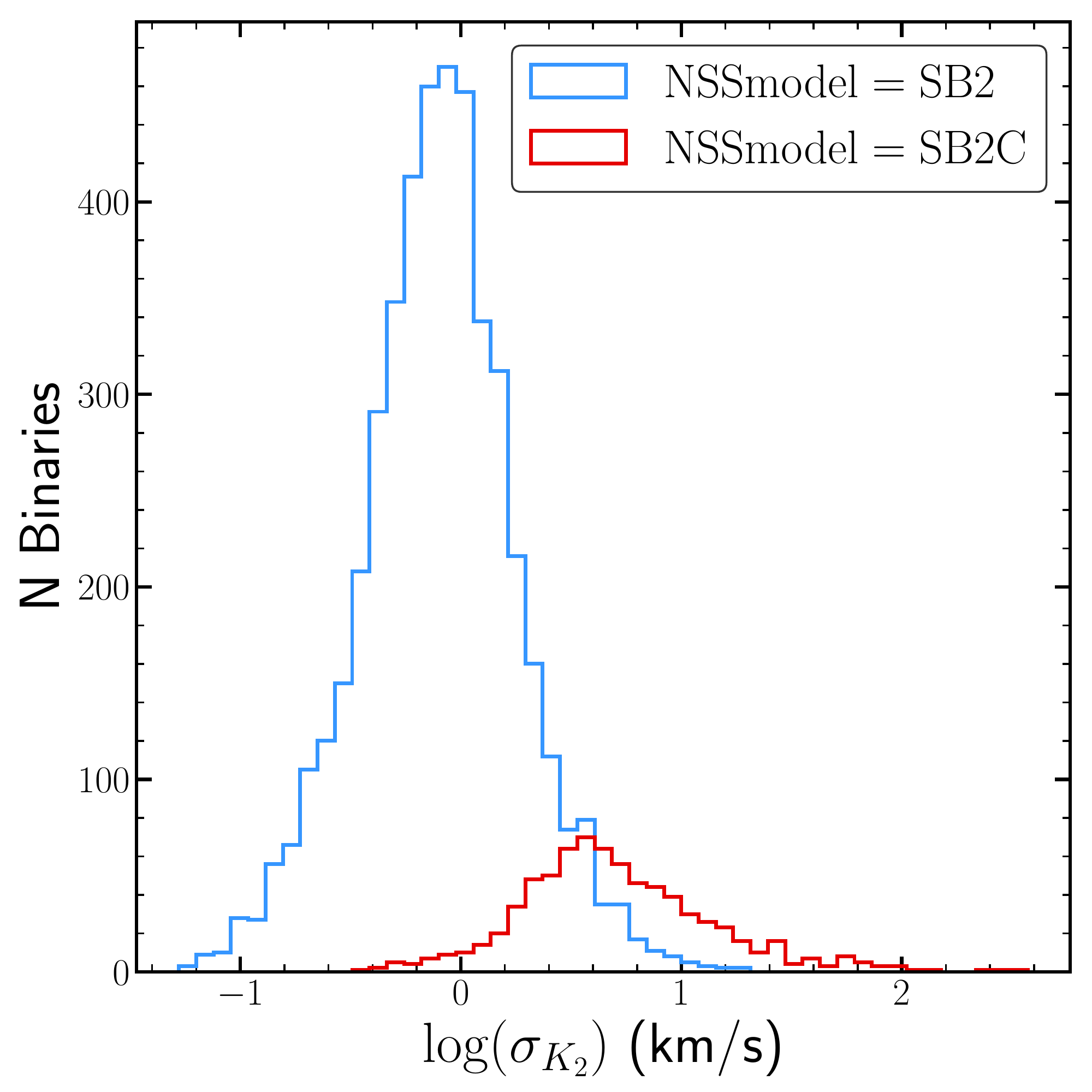}
    \caption{Distribution of the error on the velocity semi-amplitude of the secondary, $\sigma_{K_2}$ for the full sample of \Gaia{} SB2s colored by the type of model used. The \Gaia{} SB2s with the circular orbit model generally have larger errors on $K_2$.}
    \label{fig:k2_err_hist}
\end{figure}

Table \ref{tab:fp_table} reports the \Gaia{} Source information, MCMC posteriors, and the evolutionary state of the primary based its the CMD position. All of the light curves and corner plots are available online at \url{https://asas-sn.osu.edu/binaries/mass-radius}, and in the electronic version of the paper. 

Figure \ref{fig:mass_radius} shows the masses and radii of our sample. The gray dashed line shows the single star Solar metallicity zero-age main sequence (ZAMS) from the MESA Isochrones \& Stellar Tracks \citep[MIST,][]{Choi16, Dotter16}. Figure \ref{fig:mass_radius} also compares our sample to the masses and radii in \citet{Torres10}. The \citet{Torres10} catalog includes the selection criteria that both components have masses and radii errors $<3\%$, so it is not surprising that our uncertainties on the stellar parameters are larger, but we find results consistent with expectations for main-sequence stars. We report masses and radii of 12 stars on the giant branch, a sparsely populated region in the \citet{Torres10} catalog. Our catalog also has fewer high-mass main sequence stars. This is primarily due to the saturation limit of ASAS-SN, but \Gaia{} RVS data also uses the limit $T_{\rm{eff}} < 14500$~K, which is shown by the gray dotted line in Figure \ref{fig:mass_radius} for Solar-metallicity evolutionary tracks. We did identify $\sim10$ eclipsing binaries on the upper main sequence during visual inspection, but all were saturated in the ASAS-SN light curves. The lowest mass system in our catalog, CM Draconis, was also included in the \citet{Torres10} catalog, and our results are consistent within $1\sigma$. 

Some of our stars have masses and radii that place them below the ZAMS line. The deviation from the Solar metallicity ZAMS is too large to be a metallicity effect, so these stars probably have poor \Gaia{} SB2 solutions or histories of mass transfer. Some of these systems are at shorter orbital periods, but none have especially high Roche-lobe filling fractions, suggesting mass-transfer may not be the source of the deviation. These systems also do not have large \gof{} values.

There are also three stars with $M < 1$~M$_\odot$ and $R>2$~$R_\odot$. Two of these are in the same binary, \Gaia{} DR3 5339144356205089408. This system has Roche-lobe filling fractions of $f_1=0.48$ and $f_2=0.94$ for the primary and secondary, respectively, which may indicate a history of mass-transfer. The third star with low mass and large radius is \Gaia{} DR3 1762094209603752192. The other star in this binary is on the subgiant branch, with $M_1=2.15$~M$_\odot$ and $R_1=4.3$~R$_\odot$. There is also a large difference in eclipse depth (see Figure \ref{fig:mcmc_lc_176}), suggesting that the more luminous star is also hotter. Both stars have Roche-lobe filling factors $0.5 < f < 0.6$, which could suggest that if mass transfer did occur, it is not currently ongoing.

Figure \ref{fig:frac_error_hists} shows the distribution of the mass and radius uncertainties. For systems with asymmetric posteriors, we report the larger of the $\pm 1\sigma$ uncertainties. The median uncertainty on the mass is \medFracErrorMass{}\%, and the median uncertainty on the radius is \medFracErrorRadius{}\%. \Gaia{} DR3 includes two solution types for SB2s: a standard double-lined binary where the eccentricity is a free parameter ($\rm{NSSmodel}=\rm{SB2}$) and a simplified, circular model ($\rm{NSSmodel}=\rm{SB2C}$). Figure \ref{fig:frac_error_hists} shows that the binaries with the circular orbit model correspond to almost all of the systems with high fractional errors on the mass and the radius.

The SB2s with the circular orbit model also generally have higher errors on the velocity semi-amplitude of the secondary, $\sigma_{K_2}$ (Figure \ref{fig:k2_err_hist}). Since the velocity semi-amplitudes are used in our \PHOEBE{} models to set the mass ratio, projected semi-major axis, and the priors for the MCMC runs, a large $\sigma_{K_2}$ will produce a large $\sigma_M/M$. If we only consider binaries with $\rm{NSSmodel}=\rm{SB2}$, the median uncertainty on the mass is \medFracErrorMassModel{}\% and the median uncertainty on the radius is \medFracErrorRadius{}\%.

\section{Summary} \label{sec:discussion}

\Gaia{} Data Release 3 includes spectroscopic orbits for more than 181,000 single-lined spectroscopic binaries and \nSB{} double-lined spectroscopic binaries. We cross-match the catalog of SB2s with our ASAS-SN eclipsing binary catalog and further extend the sample through visual inspection. We compare the period and eccentricity from the ASAS-SN light curve to the \Gaia{} values (Figure \ref{fig:period_ecc_compare}) and find that only \fracPeriodEccMatch{}\% of systems have \Gaia{} orbits in agreement with the light curve fits. Although it is difficult to identify the source of the discrepancy in the \Gaia{} solution without the individual RVs, we do see more disagreement at shorter periods and at fainter apparent magnitudes. For the single-lined binaries, \citet{Bashi22} combined various \Gaia{} radial velocity statistics to reject poor \Gaia{} orbits and identified a sample of $\sim90,000$ of the $\sim181,000$ as having good SB1 orbits, but no such metrics currently exist for SB2s. 

For the \nPeriodEccMatch{} systems where the \Gaia{} solution is consistent with the eclipsing binary light curve, we use \PHOEBE{} to fit the light curve constrained by the velocity semi-amplitudes to derive masses and radii for \nStarsMassesRadii{} stars. Of these \nStarsMassesRadii{} stars, \nStarsMassesRadiiLowError{} have fractional mass and radius uncertainties less than 10\% (Figure \ref{fig:frac_error_hists}). We find that many of the systems with high fractional mass errors use the \Gaia{} circular orbit model ``SB2C'', which tend to have larger uncertainties on $K_2$ (Figure \ref{fig:k2_err_hist}). The uncertainties in our catalog are larger than those in the \citet{Torres10} sample of eclipsing SB2s (Figure \ref{fig:mass_radius}), but we only use publicly available data from large surveys. 

Our sample is strongly limited by the differences between the magnitude range of ASAS-SN and the \Gaia{} Radial Velocity Spectrometer. Since almost all of the \Gaia{} SB2s are bright ($G < 12$~mag), and ASAS-SN light curves start to saturate at $G\sim11$, only a fraction of the eclipsing spectroscopic binaries observed by \Gaia{} have been modeled here. Large photometric surveys of brighter stars, such as ASAS \citep{Pojmanski02, Paczynski06} and the Kilodegree Extremely Little Telescope \citep[KELT,][]{Pepper07} could be used to identify and characterize these brighter systems. The \Gaia{} light curves often have too few epochs for modeling detached eclipsing binary light curves, but future releases will provide additional $G$, $G_{\rm{BP}}$, and $G_{\rm{RP}}$ observations for millions of binaries. Multi-band light curves could also be used to constrain absolute temperatures of the stars, providing additional constraints on relations between stellar parameters. 

In total, we report masses and radii for 12 giants. One of these systems, \Gaia{} DR3 509431332327692032, has the longest period, $P=401.7$~d, and the highest eccentricity, $e=0.46$, of the binaries in our sample. Since masses of giants are difficult to determine through isochrone fitting, these systems are valuable for expanding the small sample size of giants with precise physical parameters. Long time period photometry is crucial to detecting and characterizing these systems. Our ASAS-SN eclipsing binary catalog contains more than 600 eclipsing binaries on the giant branch that could be used to expand the population of giants with dynamical masses and radii. 

The sample of spectroscopic binaries will continue to expand with future \Gaia{} data releases and and upcoming spectroscopic surveys such as Milky Way Mapper \citep{Kollmeier17}. Large catalogs of eclipsing binaries can be used not only to identify eclipsing SB2s, but also to provide initial conditions for detailed modeling of the light curve and radial velocity data. 

\section*{Acknowledgements}

We thank the \PHOEBE{} developers for their help in troubleshooting. We thank Las Cumbres Observatory and its staff for their continued support of ASAS-SN. ASAS-SN is funded in part by the Gordon and Betty Moore Foundation through grants GBMF5490 and GBMF10501 to the Ohio State University, and also funded in part by the Alfred P. Sloan Foundation grant G-2021-14192.
 
DMR, KZS, and CSK are supported by NSF grant AST-1908570. Support for TJ was provided by NASA through the NASA Hubble Fellowship grant HF2-51509 awarded by the Space Telescope Science Institute, which is operated by the Association of Universities for Research in Astronomy, Inc., for NASA, under contract NAS5-26555.

This work has made use of data from the European Space Agency (ESA)
mission {\it Gaia} (\url{https://www.cosmos.esa.int/gaia}), processed by
the {\it Gaia} Data Processing and Analysis Consortium.

\section*{Data Availability}

The ASAS-SN photometric data underlying this article are available in the ASAS-SN eclipsing
binaries database (https://asas-sn.osu.edu/binaries/mass-radius) and the ASAS-SN Photometry Database
(https://asas-sn.osu.edu/photometry). 
The data underlying this article are available in the article and in its online supplementary material.


\clearpage
\bibliographystyle{mnras}
\bibliography{asassn_gaia_sb2} 

\begin{thebibliography}{}
\makeatletter
\relax
\def\mn@urlcharsother{\let\do\@makeother \do\$\do\&\do\#\do\^\do\_\do\%\do\~}
\def\mn@doi{\begingroup\mn@urlcharsother \@ifnextchar [ {\mn@doi@}
  {\mn@doi@[]}}
\def\mn@doi@[#1]#2{\def\@tempa{#1}\ifx\@tempa\@empty \href
  {http://dx.doi.org/#2} {doi:#2}\else \href {http://dx.doi.org/#2} {#1}\fi
  \endgroup}
\def\mn@eprint#1#2{\mn@eprint@#1:#2::\@nil}
\def\mn@eprint@arXiv#1{\href {http://arxiv.org/abs/#1} {{\tt arXiv:#1}}}
\def\mn@eprint@dblp#1{\href {http://dblp.uni-trier.de/rec/bibtex/#1.xml}
  {dblp:#1}}
\def\mn@eprint@#1:#2:#3:#4\@nil{\def\@tempa {#1}\def\@tempb {#2}\def\@tempc
  {#3}\ifx \@tempc \@empty \let \@tempc \@tempb \let \@tempb \@tempa \fi \ifx
  \@tempb \@empty \def\@tempb {arXiv}\fi \@ifundefined
  {mn@eprint@\@tempb}{\@tempb:\@tempc}{\expandafter \expandafter \csname
  mn@eprint@\@tempb\endcsname \expandafter{\@tempc}}}

\bibitem[\protect\citeauthoryear{{Andersen}}{{Andersen}}{1991}]{Andersen91}
{Andersen} J.,  1991, \mn@doi [\aapr] {10.1007/BF00873538}, \href
  {https://ui.adsabs.harvard.edu/abs/1991A&ARv...3...91A} {3, 91}

\bibitem[\protect\citeauthoryear{{Bailer-Jones}, {Rybizki}, {Fouesneau},
  {Demleitner}  \& {Andrae}}{{Bailer-Jones} et~al.}{2021}]{BailerJones21}
{Bailer-Jones} C.~A.~L.,  {Rybizki} J.,  {Fouesneau} M.,  {Demleitner} M.,
  {Andrae} R.,  2021, VizieR Online Data Catalog, \href
  {https://ui.adsabs.harvard.edu/abs/2021yCat.1352....0B} {p. I/352}

\bibitem[\protect\citeauthoryear{{Bashi}, {Shahaf}, {Mazeh}, {Faigler}, {Dong},
  {El-Badry}, {Rix}  \& {Jorissen}}{{Bashi} et~al.}{2022}]{Bashi22}
{Bashi} D.,  {Shahaf} S.,  {Mazeh} T.,  {Faigler} S.,  {Dong} S.,  {El-Badry}
  K.,  {Rix} H.-W.,   {Jorissen} A.,  2022, arXiv e-prints, \href
  {https://ui.adsabs.harvard.edu/abs/2022arXiv220708832B} {p. arXiv:2207.08832}

\bibitem[\protect\citeauthoryear{{Beck} et~al.,}{{Beck} et~al.}{2014}]{Beck14}
{Beck} P.~G.,  et~al., 2014, \mn@doi [\aap] {10.1051/0004-6361/201322477},
  \href {https://ui.adsabs.harvard.edu/abs/2014A&A...564A..36B} {564, A36}

\bibitem[\protect\citeauthoryear{{Beck} et~al.,}{{Beck} et~al.}{2022}]{Beck22}
{Beck} P.~G.,  et~al., 2022, arXiv e-prints, \href
  {https://ui.adsabs.harvard.edu/abs/2022arXiv220202373B} {p. arXiv:2202.02373}

\bibitem[\protect\citeauthoryear{{Benbakoura} et~al.,}{{Benbakoura}
  et~al.}{2021}]{Benbakoura21}
{Benbakoura} M.,  et~al., 2021, \mn@doi [\aap] {10.1051/0004-6361/202037783},
  \href {https://ui.adsabs.harvard.edu/abs/2021A&A...648A.113B} {648, A113}

\bibitem[\protect\citeauthoryear{{Bhatti}, {lgbouma}  \& {Joshua}}{{Bhatti}
  et~al.}{2018}]{Bhatti18}
{Bhatti} W.,  {lgbouma}  {Joshua} 2018, {Waqasbhatti/Astrobase: Astrobase
  V0.3.8}, \mn@doi{10.5281/zenodo.1185231}

\bibitem[\protect\citeauthoryear{{B{\'o}di} \& {Hajdu}}{{B{\'o}di} \&
  {Hajdu}}{2021}]{Bodi21}
{B{\'o}di} A.,  {Hajdu} T.,  2021, \mn@doi [\apjs] {10.3847/1538-4365/ac082c},
  \href {https://ui.adsabs.harvard.edu/abs/2021ApJS..255....1B} {255, 1}

\bibitem[\protect\citeauthoryear{{Bovy}, {Rix}, {Green}, {Schlafly}  \&
  {Finkbeiner}}{{Bovy} et~al.}{2016}]{Bovy16}
{Bovy} J.,  {Rix} H.-W.,  {Green} G.~M.,  {Schlafly} E.~F.,   {Finkbeiner}
  D.~P.,  2016, \mn@doi [\apj] {10.3847/0004-637X/818/2/130}, \href
  {https://ui.adsabs.harvard.edu/abs/2016ApJ...818..130B} {818, 130}

\bibitem[\protect\citeauthoryear{{Brogaard} et~al.,}{{Brogaard}
  et~al.}{2018}]{Brogaard18}
{Brogaard} K.,  et~al., 2018, \mn@doi [\mnras] {10.1093/mnras/sty268}, \href
  {https://ui.adsabs.harvard.edu/abs/2018MNRAS.476.3729B} {476, 3729}

\bibitem[\protect\citeauthoryear{{Choi}, {Dotter}, {Conroy}, {Cantiello},
  {Paxton}  \& {Johnson}}{{Choi} et~al.}{2016}]{Choi16}
{Choi} J.,  {Dotter} A.,  {Conroy} C.,  {Cantiello} M.,  {Paxton} B.,
  {Johnson} B.~D.,  2016, \mn@doi [\apj] {10.3847/0004-637X/823/2/102}, \href
  {https://ui.adsabs.harvard.edu/abs/2016ApJ...823..102C} {823, 102}

\bibitem[\protect\citeauthoryear{{Conroy} et~al.,}{{Conroy}
  et~al.}{2020}]{Conroy20}
{Conroy} K.~E.,  et~al., 2020, \mn@doi [\apjs] {10.3847/1538-4365/abb4e2},
  \href {https://ui.adsabs.harvard.edu/abs/2020ApJS..250...34C} {250, 34}

\bibitem[\protect\citeauthoryear{{Cui} et~al.,}{{Cui} et~al.}{2012}]{Cui12}
{Cui} X.-Q.,  et~al., 2012, \mn@doi [Research in Astronomy and Astrophysics]
  {10.1088/1674-4527/12/9/003}, \href
  {https://ui.adsabs.harvard.edu/abs/2012RAA....12.1197C} {12, 1197}

\bibitem[\protect\citeauthoryear{{Dotter}}{{Dotter}}{2016}]{Dotter16}
{Dotter} A.,  2016, \mn@doi [\apjs] {10.3847/0067-0049/222/1/8}, \href
  {https://ui.adsabs.harvard.edu/abs/2016ApJS..222....8D} {222, 8}

\bibitem[\protect\citeauthoryear{{Drimmel}, {Cabrera-Lavers}  \&
  {L{\'o}pez-Corredoira}}{{Drimmel} et~al.}{2003}]{Drimmel03}
{Drimmel} R.,  {Cabrera-Lavers} A.,   {L{\'o}pez-Corredoira} M.,  2003, \mn@doi
  [\aap] {10.1051/0004-6361:20031070}, \href
  {https://ui.adsabs.harvard.edu/abs/2003A&A...409..205D} {409, 205}

\bibitem[\protect\citeauthoryear{{Duck}, {Gaudi}, {Eastman}  \&
  {Rodriguez}}{{Duck} et~al.}{2022}]{Duck22}
{Duck} A.,  {Gaudi} B.~S.,  {Eastman} J.~D.,   {Rodriguez} J.~E.,  2022, arXiv
  e-prints, \href {https://ui.adsabs.harvard.edu/abs/2022arXiv220909266D} {p.
  arXiv:2209.09266}

\bibitem[\protect\citeauthoryear{{Eastman}, {Gaudi}  \& {Agol}}{{Eastman}
  et~al.}{2013}]{Eastman13}
{Eastman} J.,  {Gaudi} B.~S.,   {Agol} E.,  2013, \mn@doi [\pasp]
  {10.1086/669497}, \href
  {https://ui.adsabs.harvard.edu/abs/2013PASP..125...83E} {125, 83}

\bibitem[\protect\citeauthoryear{{Eggleton}}{{Eggleton}}{1983}]{Eggleton83}
{Eggleton} P.~P.,  1983, \mn@doi [\apj] {10.1086/160960}, \href
  {https://ui.adsabs.harvard.edu/abs/1983ApJ...268..368E} {268, 368}

\bibitem[\protect\citeauthoryear{{Enoch}, {Collier Cameron}, {Parley}  \&
  {Hebb}}{{Enoch} et~al.}{2010}]{Enoch10}
{Enoch} B.,  {Collier Cameron} A.,  {Parley} N.~R.,   {Hebb} L.,  2010, \mn@doi
  [\aap] {10.1051/0004-6361/201014326}, \href
  {https://ui.adsabs.harvard.edu/abs/2010A&A...516A..33E} {516, A33}

\bibitem[\protect\citeauthoryear{{Foreman-Mackey}, {Hogg}, {Lang}  \&
  {Goodman}}{{Foreman-Mackey} et~al.}{2013}]{ForemanMackey13}
{Foreman-Mackey} D.,  {Hogg} D.~W.,  {Lang} D.,   {Goodman} J.,  2013, \mn@doi
  [\pasp] {10.1086/670067}, \href
  {https://ui.adsabs.harvard.edu/abs/2013PASP..125..306F} {125, 306}

\bibitem[\protect\citeauthoryear{{Frandsen} et~al.,}{{Frandsen}
  et~al.}{2013}]{Frandsen13}
{Frandsen} S.,  et~al., 2013, \mn@doi [\aap] {10.1051/0004-6361/201321817},
  \href {https://ui.adsabs.harvard.edu/abs/2013A&A...556A.138F} {556, A138}

\bibitem[\protect\citeauthoryear{{Gaia Collaboration} et~al.,}{{Gaia
  Collaboration} et~al.}{2022}]{GaiaCollaboration2022}
{Gaia Collaboration} et~al., 2022, arXiv e-prints, \href
  {https://ui.adsabs.harvard.edu/abs/2022arXiv220800211G} {p. arXiv:2208.00211}

\bibitem[\protect\citeauthoryear{{Gao} \& {Han}}{{Gao} \& {Han}}{2012}]{Gao12}
{Gao} F.,  {Han} L.,  2012, \mn@doi [Optim Appl] {10.1007/s10589-010-9329-3},
  51, 259

\bibitem[\protect\citeauthoryear{{Gaulme}, {McKeever}, {Rawls}, {Jackiewicz},
  {Mosser}  \& {Guzik}}{{Gaulme} et~al.}{2013}]{Gaulme13}
{Gaulme} P.,  {McKeever} J.,  {Rawls} M.~L.,  {Jackiewicz} J.,  {Mosser} B.,
  {Guzik} J.~A.,  2013, \mn@doi [\apj] {10.1088/0004-637X/767/1/82}, \href
  {https://ui.adsabs.harvard.edu/abs/2013ApJ...767...82G} {767, 82}

\bibitem[\protect\citeauthoryear{{Gaulme}, {Jackiewicz}, {Appourchaux}  \&
  {Mosser}}{{Gaulme} et~al.}{2014}]{Gaulme14}
{Gaulme} P.,  {Jackiewicz} J.,  {Appourchaux} T.,   {Mosser} B.,  2014, \mn@doi
  [\apj] {10.1088/0004-637X/785/1/5}, \href
  {https://ui.adsabs.harvard.edu/abs/2014ApJ...785....5G} {785, 5}

\bibitem[\protect\citeauthoryear{{Gaulme} et~al.,}{{Gaulme}
  et~al.}{2016}]{Gaulme16}
{Gaulme} P.,  et~al., 2016, \mn@doi [\apj] {10.3847/0004-637X/832/2/121}, \href
  {https://ui.adsabs.harvard.edu/abs/2016ApJ...832..121G} {832, 121}

\bibitem[\protect\citeauthoryear{{Graczyk} et~al.,}{{Graczyk}
  et~al.}{2011}]{Graczyk11}
{Graczyk} D.,  et~al., 2011, \actaa, \href
  {https://ui.adsabs.harvard.edu/abs/2011AcA....61..103G} {61, 103}

\bibitem[\protect\citeauthoryear{{Green}, {Schlafly}, {Zucker}, {Speagle}  \&
  {Finkbeiner}}{{Green} et~al.}{2019}]{Green19}
{Green} G.~M.,  {Schlafly} E.,  {Zucker} C.,  {Speagle} J.~S.,   {Finkbeiner}
  D.,  2019, \mn@doi [\apj] {10.3847/1538-4357/ab5362}, \href
  {https://ui.adsabs.harvard.edu/abs/2019ApJ...887...93G} {887, 93}

\bibitem[\protect\citeauthoryear{{Hambleton}, {Pr{\v{s}}a}, {Fleming},
  {Mahadevan}  \& {Bender}}{{Hambleton} et~al.}{2022}]{Hambleton22}
{Hambleton} K.,  {Pr{\v{s}}a} A.,  {Fleming} S.~W.,  {Mahadevan} S.,   {Bender}
  C.~F.,  2022, \mn@doi [\apj] {10.3847/1538-4357/ac69d7}, \href
  {https://ui.adsabs.harvard.edu/abs/2022ApJ...931...75H} {931, 75}

\bibitem[\protect\citeauthoryear{{Hekker} et~al.,}{{Hekker}
  et~al.}{2010}]{Hekker10}
{Hekker} S.,  et~al., 2010, \mn@doi [\apjl] {10.1088/2041-8205/713/2/L187},
  \href {https://ui.adsabs.harvard.edu/abs/2010ApJ...713L.187H} {713, L187}

\bibitem[\protect\citeauthoryear{{He{\l}miniak} et~al.,}{{He{\l}miniak}
  et~al.}{2021}]{Helminiak21}
{He{\l}miniak} K.~G.,  et~al., 2021, \mn@doi [\mnras] {10.1093/mnras/stab2963},
  \href {https://ui.adsabs.harvard.edu/abs/2021MNRAS.508.5687H} {508, 5687}

\bibitem[\protect\citeauthoryear{{Huang} et~al.,}{{Huang}
  et~al.}{2020a}]{Huang20a}
{Huang} C.~X.,  et~al., 2020a, \mn@doi [Research Notes of the American
  Astronomical Society] {10.3847/2515-5172/abca2e}, \href
  {https://ui.adsabs.harvard.edu/abs/2020RNAAS...4..204H} {4, 204}

\bibitem[\protect\citeauthoryear{{Huang} et~al.,}{{Huang}
  et~al.}{2020b}]{Huang20b}
{Huang} C.~X.,  et~al., 2020b, \mn@doi [Research Notes of the American
  Astronomical Society] {10.3847/2515-5172/abca2d}, \href
  {https://ui.adsabs.harvard.edu/abs/2020RNAAS...4..206H} {4, 206}

\bibitem[\protect\citeauthoryear{{Jayasinghe} et~al.,}{{Jayasinghe}
  et~al.}{2019}]{Jayasinghe19}
{Jayasinghe} T.,  et~al., 2019, \mn@doi [\mnras] {10.1093/mnras/stz844}, \href
  {https://ui.adsabs.harvard.edu/abs/2019MNRAS.486.1907J} {486, 1907}

\bibitem[\protect\citeauthoryear{{Jayasinghe}, {Rowan}, {Thompson}, {Kochanek}
  \& {Stanek}}{{Jayasinghe} et~al.}{2022}]{Jayasinghe22}
{Jayasinghe} T.,  {Rowan} D.~M.,  {Thompson} T.~A.,  {Kochanek} C.~S.,
  {Stanek} K.~Z.,  2022, arXiv e-prints, \href
  {https://ui.adsabs.harvard.edu/abs/2022arXiv220705086J} {p. arXiv:2207.05086}

\bibitem[\protect\citeauthoryear{{Kirk} et~al.,}{{Kirk} et~al.}{2016}]{Kirk16}
{Kirk} B.,  et~al., 2016, \mn@doi [\aj] {10.3847/0004-6256/151/3/68}, \href
  {https://ui.adsabs.harvard.edu/abs/2016AJ....151...68K} {151, 68}

\bibitem[\protect\citeauthoryear{{Kjeldsen} \& {Bedding}}{{Kjeldsen} \&
  {Bedding}}{1995}]{Kjeldsen95}
{Kjeldsen} H.,  {Bedding} T.~R.,  1995, \aap, \href
  {https://ui.adsabs.harvard.edu/abs/1995A&A...293...87K} {293, 87}

\bibitem[\protect\citeauthoryear{{Kochanek} et~al.,}{{Kochanek}
  et~al.}{2017}]{Kochanek17}
{Kochanek} C.~S.,  et~al., 2017, \mn@doi [\pasp] {10.1088/1538-3873/aa80d9},
  \href {https://ui.adsabs.harvard.edu/abs/2017PASP..129j4502K} {129, 104502}

\bibitem[\protect\citeauthoryear{{Kollmeier} et~al.,}{{Kollmeier}
  et~al.}{2017}]{Kollmeier17}
{Kollmeier} J.~A.,  et~al., 2017, arXiv e-prints, \href
  {https://ui.adsabs.harvard.edu/abs/2017arXiv171103234K} {p. arXiv:1711.03234}

\bibitem[\protect\citeauthoryear{{Kounkel} et~al.,}{{Kounkel}
  et~al.}{2021}]{Kounkel21}
{Kounkel} M.,  et~al., 2021, \mn@doi [\aj] {10.3847/1538-3881/ac1798}, \href
  {https://ui.adsabs.harvard.edu/abs/2021AJ....162..184K} {162, 184}

\bibitem[\protect\citeauthoryear{{Kov{\'a}cs}, {Zucker}  \&
  {Mazeh}}{{Kov{\'a}cs} et~al.}{2002}]{Kovacs02}
{Kov{\'a}cs} G.,  {Zucker} S.,   {Mazeh} T.,  2002, \mn@doi [\aap]
  {10.1051/0004-6361:20020802}, \href
  {https://ui.adsabs.harvard.edu/abs/2002A&A...391..369K} {391, 369}

\bibitem[\protect\citeauthoryear{{Kunimoto} et~al.,}{{Kunimoto}
  et~al.}{2021}]{Kunimoto21}
{Kunimoto} M.,  et~al., 2021, \mn@doi [Research Notes of the American
  Astronomical Society] {10.3847/2515-5172/ac2ef0}, \href
  {https://ui.adsabs.harvard.edu/abs/2021RNAAS...5..234K} {5, 234}

\bibitem[\protect\citeauthoryear{{Majewski} et~al.,}{{Majewski}
  et~al.}{2017}]{Majewski17}
{Majewski} S.~R.,  et~al., 2017, \mn@doi [\aj] {10.3847/1538-3881/aa784d},
  \href {https://ui.adsabs.harvard.edu/abs/2017AJ....154...94M} {154, 94}

\bibitem[\protect\citeauthoryear{{Marshall}, {Robin}, {Reyl{\'e}}, {Schultheis}
   \& {Picaud}}{{Marshall} et~al.}{2006}]{Marshall06}
{Marshall} D.~J.,  {Robin} A.~C.,  {Reyl{\'e}} C.,  {Schultheis} M.,   {Picaud}
  S.,  2006, \mn@doi [\aap] {10.1051/0004-6361:20053842}, \href
  {https://ui.adsabs.harvard.edu/abs/2006A&A...453..635M} {453, 635}

\bibitem[\protect\citeauthoryear{{Mowlavi} et~al.,}{{Mowlavi}
  et~al.}{2017}]{Mowlavi17}
{Mowlavi} N.,  et~al., 2017, \mn@doi [\aap] {10.1051/0004-6361/201730613},
  \href {https://ui.adsabs.harvard.edu/abs/2017A&A...606A..92M} {606, A92}

\bibitem[\protect\citeauthoryear{{Paczy{\'n}ski}, {Szczygie{\l}}, {Pilecki}  \&
  {Pojma{\'n}ski}}{{Paczy{\'n}ski} et~al.}{2006}]{Paczynski06}
{Paczy{\'n}ski} B.,  {Szczygie{\l}} D.~M.,  {Pilecki} B.,   {Pojma{\'n}ski} G.,
   2006, \mn@doi [\mnras] {10.1111/j.1365-2966.2006.10223.x}, \href
  {https://ui.adsabs.harvard.edu/abs/2006MNRAS.368.1311P} {368, 1311}

\bibitem[\protect\citeauthoryear{{Pawlak} et~al.,}{{Pawlak}
  et~al.}{2013}]{Pawlak13}
{Pawlak} M.,  et~al., 2013, \actaa, \href
  {https://ui.adsabs.harvard.edu/abs/2013AcA....63..323P} {63, 323}

\bibitem[\protect\citeauthoryear{{Pepper} et~al.,}{{Pepper}
  et~al.}{2007}]{Pepper07}
{Pepper} J.,  et~al., 2007, \mn@doi [\pasp] {10.1086/521836}, \href
  {https://ui.adsabs.harvard.edu/abs/2007PASP..119..923P} {119, 923}

\bibitem[\protect\citeauthoryear{{Petrosky}, {Hwang}, {Zakamska}, {Chandra}  \&
  {Hill}}{{Petrosky} et~al.}{2021}]{Petrosky21}
{Petrosky} E.,  {Hwang} H.-C.,  {Zakamska} N.~L.,  {Chandra} V.,   {Hill}
  M.~J.,  2021, \mn@doi [\mnras] {10.1093/mnras/stab592}, \href
  {https://ui.adsabs.harvard.edu/abs/2021MNRAS.503.3975P} {503, 3975}

\bibitem[\protect\citeauthoryear{{Pietrukowicz} et~al.,}{{Pietrukowicz}
  et~al.}{2013}]{Pietrukowicz13}
{Pietrukowicz} P.,  et~al., 2013, \actaa, \href
  {https://ui.adsabs.harvard.edu/abs/2013AcA....63..115P} {63, 115}

\bibitem[\protect\citeauthoryear{{Pojmanski}}{{Pojmanski}}{2002}]{Pojmanski02}
{Pojmanski} G.,  2002, \actaa, \href
  {https://ui.adsabs.harvard.edu/abs/2002AcA....52..397P} {52, 397}

\bibitem[\protect\citeauthoryear{{Pourbaix} et~al.,}{{Pourbaix}
  et~al.}{2004}]{Pourbaix04}
{Pourbaix} D.,  et~al., 2004, \mn@doi [\aap] {10.1051/0004-6361:20041213},
  \href {https://ui.adsabs.harvard.edu/abs/2004A&A...424..727P} {424, 727}

\bibitem[\protect\citeauthoryear{{Pr{\v{s}}a} \& {Zwitter}}{{Pr{\v{s}}a} \&
  {Zwitter}}{2005}]{Prsa05}
{Pr{\v{s}}a} A.,  {Zwitter} T.,  2005, \mn@doi [\apj] {10.1086/430591}, \href
  {https://ui.adsabs.harvard.edu/abs/2005ApJ...628..426P} {628, 426}

\bibitem[\protect\citeauthoryear{{Pr{\v{s}}a}, {Guinan}, {Devinney},
  {DeGeorge}, {Bradstreet}, {Giammarco}, {Alcock}  \& {Engle}}{{Pr{\v{s}}a}
  et~al.}{2008}]{Prsa08}
{Pr{\v{s}}a} A.,  {Guinan} E.~F.,  {Devinney} E.~J.,  {DeGeorge} M.,
  {Bradstreet} D.~H.,  {Giammarco} J.~M.,  {Alcock} C.~R.,   {Engle} S.~G.,
  2008, \mn@doi [\apj] {10.1086/591783}, \href
  {https://ui.adsabs.harvard.edu/abs/2008ApJ...687..542P} {687, 542}

\bibitem[\protect\citeauthoryear{{Pr{\v{s}}a} et~al.,}{{Pr{\v{s}}a}
  et~al.}{2011}]{Prsa11}
{Pr{\v{s}}a} A.,  et~al., 2011, \mn@doi [\aj] {10.1088/0004-6256/141/3/83},
  \href {https://ui.adsabs.harvard.edu/abs/2011AJ....141...83P} {141, 83}

\bibitem[\protect\citeauthoryear{{Pr{\v{s}}a} et~al.,}{{Pr{\v{s}}a}
  et~al.}{2016}]{Prsa16}
{Pr{\v{s}}a} A.,  et~al., 2016, \mn@doi [\apjs] {10.3847/1538-4365/227/2/29},
  \href {https://ui.adsabs.harvard.edu/abs/2016ApJS..227...29P} {227, 29}

\bibitem[\protect\citeauthoryear{{Pr{\v{s}}a} et~al.,}{{Pr{\v{s}}a}
  et~al.}{2022}]{Prsa22}
{Pr{\v{s}}a} A.,  et~al., 2022, \mn@doi [\apjs] {10.3847/1538-4365/ac324a},
  \href {https://ui.adsabs.harvard.edu/abs/2022ApJS..258...16P} {258, 16}

\bibitem[\protect\citeauthoryear{{Qian}, {He}, {Zhang}, {Zhu}, {Shi}, {Zhao}
  \& {Zhou}}{{Qian} et~al.}{2017}]{Qian17}
{Qian} S.-B.,  {He} J.-J.,  {Zhang} J.,  {Zhu} L.-Y.,  {Shi} X.-D.,  {Zhao}
  E.-G.,   {Zhou} X.,  2017, \mn@doi [Research in Astronomy and Astrophysics]
  {10.1088/1674-4527/17/8/87}, \href
  {https://ui.adsabs.harvard.edu/abs/2017RAA....17...87Q} {17, 087}

\bibitem[\protect\citeauthoryear{{Qian}, {Zhang}, {He}, {Zhu}, {Zhao}, {Shi},
  {Zhou}  \& {Han}}{{Qian} et~al.}{2018}]{Qian18}
{Qian} S.~B.,  {Zhang} J.,  {He} J.~J.,  {Zhu} L.~Y.,  {Zhao} E.~G.,  {Shi}
  X.~D.,  {Zhou} X.,   {Han} Z.~T.,  2018, \mn@doi [\apjs]
  {10.3847/1538-4365/aaa601}, \href
  {https://ui.adsabs.harvard.edu/abs/2018ApJS..235....5Q} {235, 5}

\bibitem[\protect\citeauthoryear{{Ratajczak} et~al.,}{{Ratajczak}
  et~al.}{2021}]{Ratajczak21}
{Ratajczak} M.,  et~al., 2021, \mn@doi [\mnras] {10.1093/mnras/staa3488}, \href
  {https://ui.adsabs.harvard.edu/abs/2021MNRAS.500.4972R} {500, 4972}

\bibitem[\protect\citeauthoryear{{Ricker} et~al.,}{{Ricker}
  et~al.}{2015}]{Ricker15}
{Ricker} G.~R.,  et~al., 2015, \mn@doi [Journal of Astronomical Telescopes,
  Instruments, and Systems] {10.1117/1.JATIS.1.1.014003}, \href
  {https://ui.adsabs.harvard.edu/abs/2015JATIS...1a4003R} {1, 014003}

\bibitem[\protect\citeauthoryear{{Rodr{\'\i}guez Mart{\'\i}nez}
  et~al.,}{{Rodr{\'\i}guez Mart{\'\i}nez} et~al.}{2022}]{Rodriguez22}
{Rodr{\'\i}guez Mart{\'\i}nez} R.,  et~al., 2022, arXiv e-prints, \href
  {https://ui.adsabs.harvard.edu/abs/2022arXiv220807883R} {p. arXiv:2208.07883}

\bibitem[\protect\citeauthoryear{{Rowan} et~al.,}{{Rowan}
  et~al.}{2022a}]{Rowan22II}
{Rowan} D.~M.,  et~al., 2022a, arXiv e-prints, \href
  {https://ui.adsabs.harvard.edu/abs/2022arXiv221006486R} {p. arXiv:2210.06486}

\bibitem[\protect\citeauthoryear{{Rowan} et~al.,}{{Rowan}
  et~al.}{2022b}]{Rowan22}
{Rowan} D.~M.,  et~al., 2022b, \mn@doi [\mnras] {10.1093/mnras/stac2520}, \href
  {https://ui.adsabs.harvard.edu/abs/2022MNRAS.517.2190R} {517, 2190}

\bibitem[\protect\citeauthoryear{{Shappee} et~al.,}{{Shappee}
  et~al.}{2014}]{Shappee14}
{Shappee} B.~J.,  et~al., 2014, \mn@doi [\apj] {10.1088/0004-637X/788/1/48},
  \href {https://ui.adsabs.harvard.edu/abs/2014ApJ...788...48S} {788, 48}

\bibitem[\protect\citeauthoryear{{Slawson} et~al.,}{{Slawson}
  et~al.}{2011}]{Slawson11}
{Slawson} R.~W.,  et~al., 2011, \mn@doi [\aj] {10.1088/0004-6256/142/5/160},
  \href {https://ui.adsabs.harvard.edu/abs/2011AJ....142..160S} {142, 160}

\bibitem[\protect\citeauthoryear{{Soszy{\'n}ski} et~al.,}{{Soszy{\'n}ski}
  et~al.}{2016}]{Soszynski16}
{Soszy{\'n}ski} I.,  et~al., 2016, \actaa, \href
  {https://ui.adsabs.harvard.edu/abs/2016AcA....66..405S} {66, 405}

\bibitem[\protect\citeauthoryear{{Steinmetz} et~al.,}{{Steinmetz}
  et~al.}{2006}]{Steinmentz06}
{Steinmetz} M.,  et~al., 2006, \mn@doi [\aj] {10.1086/506564}, \href
  {https://ui.adsabs.harvard.edu/abs/2006AJ....132.1645S} {132, 1645}

\bibitem[\protect\citeauthoryear{{Theme{\ss}l} et~al.,}{{Theme{\ss}l}
  et~al.}{2018}]{Themebl18}
{Theme{\ss}l} N.,  et~al., 2018, \mn@doi [\mnras] {10.1093/mnras/sty1113},
  \href {https://ui.adsabs.harvard.edu/abs/2018MNRAS.478.4669T} {478, 4669}

\bibitem[\protect\citeauthoryear{{Torres}, {Andersen}  \&
  {Gim{\'e}nez}}{{Torres} et~al.}{2010}]{Torres10}
{Torres} G.,  {Andersen} J.,   {Gim{\'e}nez} A.,  2010, \mn@doi [\aapr]
  {10.1007/s00159-009-0025-1}, \href
  {https://ui.adsabs.harvard.edu/abs/2010A&ARv..18...67T} {18, 67}

\makeatother
\end{thebibliography}





\bsp	
\label{lastpage}
\end{document}